\documentclass[a4paper]{panl}
\usepackage{cite}
\usepackage{wrapfig}
\usepackage{graphicx}
\usepackage{amssymb}
\usepackage{amsfonts}
\usepackage{amsmath}
\usepackage{longtable}
\usepackage{rotating}
\usepackage{lscape}
\usepackage{epsfig}
\usepackage{multirow}
\usepackage{physics}
\usepackage{subcaption}

\originalTeX
\begin{document}
% Journal sections (see http://pkp.jinr.ru/index.php/PEPAN_LETTERS/about/editorialPolicies#focusAndScope)
%\issuearea{Physics of Elementary Particles and Atomic Nuclei. Theory}

\title{\vspace{-2cm}
	\begin{flushright}
		{\normalsize INR-TH-2024-014}
	\end{flushright}
	\vspace{0.5cm} Dark photon production via inelastic proton bremsstrahlung with Pauli form factor}
\maketitle
\authors{D.\,Gorbunov$^{a,b,}$\footnote{E-mail: gorby@inr.ac.ru},
E.\,Kriukova$^{a,c}$\footnote{E-mail: kryukova.ea15@physics.msu.ru}}
\setcounter{footnote}{0}
\from{$^{a}$\,Institute for Nuclear Research of the Russian Academy of Sciences, 60th October Anniversary pr-ct 7a, Moscow 117312, Russia}
\from{$^{b}$\,Landau Phystech School of Physics and Research, Moscow Institute of Physics and Technology,
Institutskiy per. 9, Dolgoprudny 141700, Russia}
\from{$^{c}$\,Faculty of Physics, Lomonosov Moscow State University, Leninskiye Gory 1-2, Moscow 119991, Russia}

\begin{abstract}

We study the production of vector portal mediators, dark photons, with masses in the range 0.4--1.8\,GeV in proton-proton collisions via the process of inelastic proton bremsstrahlung. In contrast to previous studies, we take into account the contribution of Pauli electromagnetic form factor to differential cross section and introduce two new splitting functions of the proton. We show that their contributions can become leading in the mass region 0.9--1.8\,GeV and present the updated estimate for the sensitivity of the future SHiP experiment to visible decays of dark photons.
\end{abstract}
\vspace*{6pt}


\section{Introduction}
\label{sec:intro}

In recent years, the portal formalism has been intensively used to describe the structure of the theories that extend the Standard Model (SM)~\cite{Lanfranchi:2020crw}. The main idea is to introduce the special particles called mediators that can feebly interact both with the SM particles and with the hidden sector. In this paper we study the phenomenology of the so-called dark photons --- the massive mediators of the vector portal. 

We consider the minimal version of dark photon model, where it kinetically mixes with the SM photon and thus interacts only with the SM particles with non-zero electric charge. The Lagrangian is the following~\cite{Okun:1982xi}
\begin{equation}
    \mathcal{L}=\mathcal{L}_\text{SM}-\frac{1}{4}F^\prime_{\mu \nu}F^{\prime \mu \nu}+\frac{\epsilon}{2}F^\prime_{\mu\nu}B^{\mu \nu}+\frac{m_{\gamma'}^2}{2}A^\prime_\mu A^{\prime \mu},
\end{equation}
where $\mathcal{L}_\text{SM}$ is the SM Lagrangian, $A^{\prime \mu}$ is the dark photon field, $F^{\prime \mu \nu}$ and $B^{\mu \nu}$ are the field strength tensors of dark photon and $U(1)_Y$ gauge field of the SM, $\epsilon$ is the kinetic mixing parameter and $m_{\gamma^\prime}$ is the dark photon mass.

The searches for minimal dark photons have been performed by many experiments~\cite{Raggi:2015yfk}, but the parameter space $(m_{\gamma^\prime}, \epsilon)$ is still constrained weakly for their masses about $\mathcal{O}(1)$~GeV. It is planned to search for such dark photons in the future by considering their potential production in $pp$-collisions in the neutrino experiments DUNE (Fermilab~\cite{Berryman:2019dme}) and T2K (J-PARC~\cite{Araki:2023xgb}) and also in the fixed-target experiment SHiP at CERN SPS~\cite{SHiP:2020vbd}. Therefore it is important to make reliable predictions for dark photon production cross sections in $pp$-collisions and its decay width at the intermediate mass scale about $\mathcal{O}(1)$~GeV, where neither QCD, nor chiral perturbation theory work properly.

The dominant production mechanism is determined by dark photon mass $m_{\gamma^\prime}$~\cite{Miller:2021ycl}. Dark photons with masses below 0.4\,GeV are mostly produced in the decays of $\pi^0$- and $\eta$-mesons due the mixing with the SM photon. Heavier dark photons with masses above  1.8\,GeV could be produced mainly in the analog of QCD Drell-Yan process. In this paper we consider inelastic proton bremsstrahlung, which is the most important production channel for dark photon masses in the intermediate region 0.4--1.8\,GeV. During the initial state radiation in the inelastic bremsstrahlung the incident proton emits the SM photon that mixes with the dark photon. The SM photon-proton-proton vertex has to be described with the electromagnetic proton form factors, that play a crucial role in this study.

The matrix element of the electromagnetic current $j^{em}_\mu \equiv \sum \bar{q} Q \gamma_\mu q$ for light quarks $q=\{u,d,s\}$ with electric charges $Q=\{2/3,-1/3,-1/3\}$ describing the creation of the nucleon-antinucleon pair $N\overline{N}$ with the momenta $p_1$ and $p_2$
\begin{equation}
    J_\mu\equiv\bra{N(p_1)\overline{N}(p_2)}j^{em}_\mu(0)\ket{0}
\end{equation}
can be parameterized with the help of Dirac $F^N_1(t)$ и Pauli $F^N_2(t)$ nucleon electromagnetic form factors as  functions of the Mandelstam variable $t \equiv (p_1+p_2)^2$ \cite{Lin:2021umz}
\begin{equation}
    J^\mu = \overline{u}(p_1) \left[F^N_1(t)\gamma_\mu + i\frac{F^N_2(t)}{2m}\sigma_{\mu\nu}(p_1^\nu+p_2^\nu)\right] v(p_2).
\end{equation}
Depending on the sign of $t$, the electromagnetic form factors are measured by two kinds of experiments. In the space-like region with $t<0$ they are studied in the process of electron-nucleon scattering $eN\rightarrow eN$ in a big number of experiments among which are PRad, Jefferson Lab and MAMI~\cite{Lin:2021xrc}. In this region the measurement of the differential scattering cross section allows to determine the real-valued Sachs form factors 
\begin{align}
    G_E(t)&\equiv F_1(t)+\frac{t}{4M^2} F_2(t),\\
    G_M(t)&\equiv F_1(t)+F_2(t)
\end{align}
and therefore the electromagnetic nucleon form factors in the space-like region, $t<0$, are known with good accuracy for the processes with small $\sqrt{-t}$. In the time-like region BABAR, BESIII, CMD-3, SND experiments etc. study the annihilation process $e^+ e^-\rightarrow N\overline{N}$ above the threshold, $\sqrt{t}>2M_N$, where $M_N$ is the nucleon mass~\cite{Lin:2021xrc}. In the time-like region, $t>0$, the form factors become complex-valued, so it is possible to measure only $|G_E(t)/G_M(t)|$ and the effective form factor 	
\begin{equation}
    |G_\text{eff}(t)|\equiv \sqrt{\frac{|G_E(t)|^2+t/2M_N^2|G_M(t)|^2}{1+t/2M_N^2}},
\end{equation}
which leads to greater uncertainties than in the space-like region. Finally, there is also the so-called \textit{unphysical} region for the time-like $\sqrt{t}<2M_N$ that cannot be studied in the current experiments mentioned above.

In this work we  use the values of electromagnetic form factors at $t=m^2_{\gamma^\prime}$ relevant for producing an on-shell massive dark photon in the process of inelastic proton bremsstrahlung. We consider the dark photon masses in the region 0.4--1.8\,GeV, and hence are unable to rely on the experimental data for this problem. Instead, we have to use the extrapolation of the fits to the unphysical region that in general can be obtained in numerous ways, for instance, by introducing the fictitious resonances in the framework of the extended vector meson dominance model~\cite{Faessler:2009tn}, or by unitary and analytic continuation of the model based on the existing resonances with quantum numbers $J^{PC}=1^{--}$ (corresponding to the SM photon)~\cite{Adamuscin:2016rer}, or in the dispersion relations approach~\cite{Lin:2021xrc}. Here we conservatively use the results presented in~\cite{Faessler:2009tn}, which is a common choice for the studies of dark photon phenomenology.

The paper is organized as follows. In section~\ref{sec:inelastic-calculation} we revisit the calculation of the inelastic bremsstrahlung differential cross section paying special attention to both Dirac $F_1(t)$ and Pauli $F_2(t)$ electromagnetic form factors of proton. Section~\ref{sec:full-cr-sec} contains the numerical results on the full inelastic bremsstrahlung cross section and the discussion of impacts of the contributions proportional to different quadratic combinations of the form factors. In order to show the importance of the new contributions, we also present a new estimate for the sensitivity of SHiP experiment to dark photons in section~\ref{sec:sensitivity}. Finally, section~\ref{sec:conclude} sums up our findings and describes perspectives for the future work.

\section{Splitting functions for inelastic bremsstrahlung including both Dirac and Pauli form factors}
\label{sec:inelastic-calculation}
In this section we study the initial state radiation of dark photon in the process of inelastic proton bremsstrahlung, see Fig.\,\ref{fig:Feynman}. 
%============================= Fig. 1 ================================
\begin{figure}[t]
\begin{center}
\includegraphics[width=0.3\textwidth]{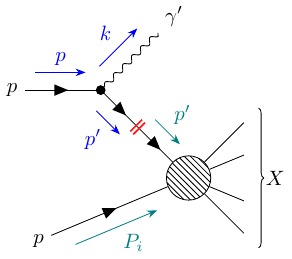}
\vspace{-3mm}
\caption{Feynman diagram for the dark photon production via inelastic proton bremsstrahlung.}
\end{center}
\labelf{fig:Feynman}
\vspace{-5mm}
\end{figure}
%============================= Fig. 1 ================================
Our final goal is to factorize the cross section of the process $pp\rightarrow \gamma^\prime X$ using the method, initially formulated in \cite{Altarelli:1977zs} for massless particles and later adjusted for the production of massive scalars \cite{Boiarska:2019jym} and for proton bremsstrahlung of dark photons \cite{Foroughi-Abari:2021zbm}. Although the calculation in \cite{Foroughi-Abari:2021zbm} is correct and is presented clearly, we argue that it does not give the \textit{final} answer for the proton bremsstrahlung cross section, since it completely ignores the contribution to the matrix element proportional to Pauli electromagnetic proton form factor $F_2(t)$. Here we extend the approach of Ref.\,\cite{Foroughi-Abari:2021zbm} and account for both Dirac $F_1(t)$ and Pauli $F_2(t)$ electromagnetic form factors. 

The particles momenta indicated in Fig.\,\ref{fig:Feynman} are $p$ for the incident proton, $P_i$ for the target proton, $k$ for the dark photon and $p^\prime$ for virtual proton. Unlike the approach of \cite{Altarelli:1977zs, Boiarska:2019jym}, we do \textit{not} use the old-fashioned perturbation theory in the calculation. Thus the momentum is conserved in every vertex and $p^\prime\equiv p-k$. The four-momenta components of two other particles involved in the dark photon emission in the Cartesian coordinate system with $z$-axis oriented along the incident proton beam are
\begin{align}
    p &= \{E_p, 0, 0, P\},\quad E_p\equiv P + \frac{M^2}{2P},\\
    k &= \{E_k, k_\perp \cos \varphi, k_\perp \sin \varphi, zP\},\quad E_k\equiv zP + \frac{m^2_{\gamma^\prime}+k^2_\perp}{2zP},
\end{align}
where $M$ is the proton mass, $P$ is the value of incident proton 3-momentum, $z$ is the fraction of 3-momentum that is taken by dark photon from the incident proton in $z$-direction, $k_\perp$ is the transverse part of dark photon momentum and hereafter we assume that $M^2\ll P^2$ and $m^2_{\gamma^\prime}+k^2_\perp \ll z^2P^2$.

We would like to present the differential cross section as the product of the splitting function $w(z, k^2_\perp)$ that describes the probability of dark photon emission and the total cross section of inelastic $pp$-scattering
\begin{equation} \label{eq:diff-cr-sec}
    \frac{\dd^2 \sigma(pp\rightarrow \gamma^\prime X)}{\dd z \dd k^2_\perp}\simeq w(z, k^2_\perp) \sigma(pp\rightarrow X).
\end{equation}
In order to do so we decompose the proton propagator numerator as the sum over proton polarizations $r^\prime$
\begin{equation}
    \hat{p}-\hat{k}+M = \sum_{r^\prime} u^{r^\prime}(p-k)\overline{u}^{r^\prime}(p-k)
\end{equation}
and introduce two vertex functions
\begin{align}
    V_1^{r^\prime r \lambda} &\equiv \overline{u}^{r^\prime}(p-k) \widehat{(\epsilon^\lambda)^*} u^r(p),\\
    V_2^{r^\prime r \lambda} &\equiv \frac{1}{4M} \overline{u}^{r^\prime}(p-k) \left[\widehat{(\epsilon^\lambda)^*}, \hat{k}\right] u^r(p)
\end{align}
also depending on the polarization of incident proton $r$ and dark photon $\lambda$. It allows us to extract the inelastic scattering subprocess amplitude $\mathcal{M}^{r^\prime}_{pp\rightarrow X}$ from the inelastic bremsstrahlung amplitude $\mathcal{M}^{r\lambda}_{pp\rightarrow \gamma^\prime X}$ giving 
\begin{equation} \label{eq:amplitude}
    \mathcal{M}^{r\lambda}_{pp\rightarrow \gamma^\prime X}=-\sum_{r^\prime} \mathcal{M}^{r^\prime}_{pp\rightarrow X} \frac{\epsilon ez}{H(z, k^2_\perp)} \left(V_1^{r^\prime r \lambda} F_1\left(m^2_{\gamma^\prime}\right)+V_2^{r^\prime r \lambda} F_2\left(m^2_{\gamma^\prime}\right)\right),
\end{equation}
where $e$ is the proton charge and the kinematic combination $H(z, k^2_\perp)\equiv k^2_\perp+(1-z)m^2_{\gamma^\prime}+z^2M^2$ represents the virtuality of intermediate proton, 
\begin{equation} \label{eq:virtuality}
    (p-k)^2=M^2-H/z.
\end{equation}
We would like to stress that \eqref{eq:amplitude} contains not only the term proportional to Dirac form factor $F_1(m^2_{\gamma^\prime})$, earlier considered by \cite{Foroughi-Abari:2021zbm}, but also the second novel term proportional to Pauli form factor $F_2(m^2_{\gamma^\prime})$.

Keeping in mind the chain of inequalities for values of Dirac and Pauli electromagnetic form factors taken at $t=m^2_{\gamma^\prime}$,  
\begin{equation}
    |F_2|^2 > |F_1F^*_2+F_2F^*_1| > |F_1|^2 > |F_1F^*_2-F_2F^*_1|,
\end{equation}
that holds for the most part of the considered dark photon mass region, one can square the amplitude \eqref{eq:amplitude} and neglect the term proportional to $\left(F_1F^*_2-F_2F^*_1\right)$ to obtain the differential cross section of inelastic bremsstrahlung in the form \eqref{eq:diff-cr-sec} with the master splitting function
\begin{equation} \label{eq:splitfunc}
    w(z, k^2_\perp)\equiv w_{11}(z, k^2_\perp)|F_1|^2+w_{22}(z, k^2_\perp)|F_2|^2+w_{12}(z, k^2_\perp)\left(F_1F^*_2+F_2F^*_1\right)
\end{equation}
and three auxiliary splitting functions that do not depend on the electromagnetic form factors
\begin{align} \label{eq:auxiliary1}
    w_{11}(z, k^2_\perp)&\equiv \frac{\epsilon^2 \alpha_{em}}{2\pi H(z, k^2_\perp)} \left(z-\frac{z\left(1-z\right)}{H(z, k^2_\perp)}\left(2M^2+m^2_{\gamma^\prime}\right)+\frac{H(z, k^2_\perp)}{2zm^2_{\gamma^\prime}}\right),\\
    w_{22}(z, k^2_\perp)&\equiv \frac{\epsilon^2 \alpha_{em}}{2\pi H} \frac{m^2_{\gamma^\prime}}{8M^2} \left(z-\frac{z\left(1-z\right)}{H(z, k^2_\perp)}\left(8M^2+m^2_{\gamma^\prime}\right)+\frac{2H(z, k^2_\perp)}{zm^2_{\gamma^\prime}}\right),\\ \label{eq:auxiliary2}
    w_{12}(z, k^2_\perp)&\equiv \frac{\epsilon^2 \alpha_{em}}{2\pi H(z, k^2_\perp)} \left(\frac{3z}{4}-\frac{3m^2_{\gamma^\prime}z\left(1-z\right)}{2H(z, k^2_\perp)}\right),
\end{align}
where $\alpha_{em}$ is the fine-structure constant. The auxiliary splitting function $w_{11}(z, k^2_\perp)$ agrees with \cite{Foroughi-Abari:2021zbm}. We compare the contributions to the total cross section of each of the auxiliary functions in the next section.

Finally, we include the off-shell hadronic form factor with the scale $\Lambda=1.5$ GeV \cite{Feuster:1998cj}
\begin{equation}
    F_\text{virt}\equiv \frac{\Lambda^4}{\Lambda^4+H^2(z, k^2_\perp)/z^2}
\end{equation}
that suppresses the amplitude for intermediate protons which are far from the mass shell (see eq.~\eqref{eq:virtuality}) and use the following fit for non-single diffractive cross section \cite{Likhoded:2010pc}
\begin{equation}
    \sigma_\text{NSD}(s)=1.76+19.8\left(\frac{s}{\text{GeV}^2}\right)^{0.057}\text{ mb},
\end{equation}
that we take at the squared center-of-mass energy of protons with momenta $p^\prime$ and $P_i$,
\begin{equation}
    \bar s(z, k^2_\perp) = 2 MP(1 - z) + 2 M^2 - H(z, k^2_\perp)/z.
\end{equation}
Thus we obtain the differential inelastic bremsstrahlung cross section, that we use further for numerical estimates 
\begin{equation} \label{eq:final}
    \frac{\dd^2 \sigma(pp\rightarrow \gamma^\prime X)}{\dd z \dd k^2_\perp}\simeq w(z, k^2_\perp) F^2_\text{virt}(z, k^2_\perp)\sigma_\text{NSD}(\bar s(z, k^2_\perp)).
\end{equation}

\section{Full inelastic bremsstrahlung cross section}
\label{sec:full-cr-sec}
\begin{figure}[t]
	\begin{center}
		\begin{subfigure}{0.49\textwidth}
			\centering
			\includegraphics[width=\textwidth]{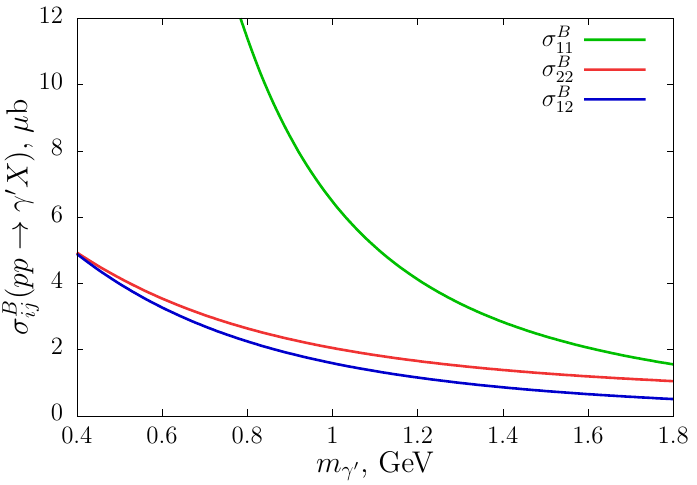}
                \vspace{-3mm}
			\caption{}
			\label{fig:bare-cr-sec}
		\end{subfigure}
		\hfill
		\begin{subfigure}{0.49\textwidth}
			\centering
			\includegraphics[width=\textwidth]{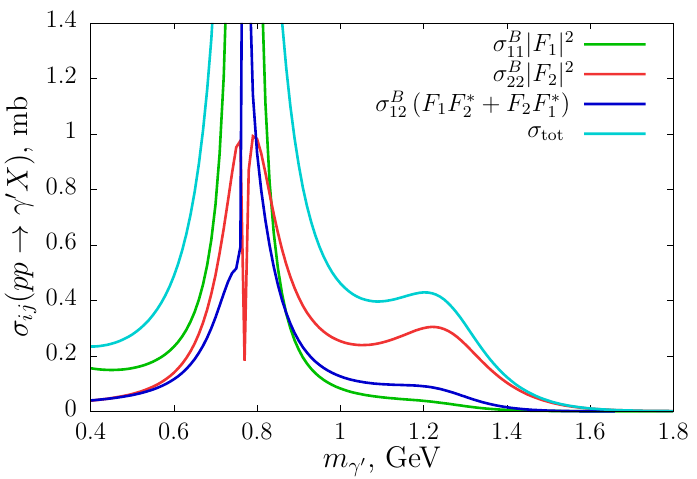}
                \vspace{-3mm}
			\caption{}
			\label{fig:full-cr-sec}
		\end{subfigure}
	\end{center}
        \vspace{-3mm}
	\caption{Integrated cross sections of inelastic proton bremsstrahlung: (a) ``bare'' cross sections \eqref{eq:bare}, (b) full cross sections including Dirac and Pauli form factors. Both plots contain lines corresponding to each of three auxiliary splitting functions \eqref{eq:auxiliary1}--\eqref{eq:auxiliary2} and are given as functions of dark photon mass $m_{\gamma'}$. Light blue line in Fig.\,\ref{fig:full-cr-sec} shows the total integrated cross section. The momentum of incident proton beam in the lab frame is $P=120\text{ GeV}$.}
\vspace{-5mm}
\end{figure}

In order to understand, whether new terms in \eqref{eq:splitfunc} are significant or negligible, in Fig.\,\ref{fig:bare-cr-sec}
%============================= Fig. 2 ================================
%============================= Fig. 2 ================================
we compare various integrated ``bare'' inelastic bremsstrahlung cross sections
\begin{equation} \label{eq:bare}
    \sigma^B_{ij}(pp\rightarrow \gamma^\prime X)\equiv\int w_{ij}(z, k^2_\perp) F^2_\text{virt}(z, k^2_\perp)\sigma_\text{NSD}(\bar s(z, k^2_\perp)) \dd z \dd k^2_\perp,
\end{equation}
where we take the momentum of incident proton beam in the lab frame $P=120\text{ GeV}$ and {\it for the moment} put all electromagnetic form factors equal to unity. For example, one can suppose that $F_1(m^2_{\gamma^\prime})\simeq F_2(m^2_{\gamma^\prime})$ and we have put them out of brackets. Although we know that it is not true, this assumption allows us to compare those parts of cross section terms which do not depend on electromagnetic proton form factors. As one can see from Fig.\,\ref{fig:bare-cr-sec}, the cross section $\sigma^B_{11}(pp\rightarrow \gamma^\prime X)$ that should be multiplied by the square of Dirac form factor, exceeds two other cross sections $\sigma^B_{22}(pp\rightarrow \gamma^\prime X)$ and $\sigma^B_{12}(pp\rightarrow \gamma^\prime X)$ connected with Pauli form factor for all considered values of dark photon mass $m_{\gamma'}$, which explains why it seemed natural to neglect these terms earlier.

Next, we restore the values of proton electromagnetic form factors $F_1(m^2_{\gamma^\prime})$ and $F_2(m^2_{\gamma^\prime})$ using the fit \cite{Faessler:2009tn} and compare the resulting integrated cross sections in Fig.\,\ref{fig:full-cr-sec}. The total inelastic bremsstrahlung cross section
\begin{equation} \label{eq:total}
    \sigma_\text{tot}=\sigma^B_{11}|F_1|^2+\sigma^B_{22}|F_2|^2+\sigma^B_{12}\left(F_1F^*_2+F_2F^*_1\right)
\end{equation}
is shown in light blue. It can be seen that in the region of dark photon masses from 0.9 to 1.8 GeV the major contribution to total cross section comes from the term proportional to the square of Pauli form factor $|F_2|^2$. It is evident from Fig.~\ref{fig:bare-cr-sec}, that this feature is due to the increase of the Pauli electromagnetic form factor in comparison with the Dirac ones. To exclude the possibility that this happens only for the chosen fit~\cite{Faessler:2009tn}, we have also studied the same quantities using another fit from~\cite{Adamuscin:2016rer}, that was obtained in completely different way. We find that the qualitative part of results remains the same. 

\section{SHiP sensitivity to dark photons}
\label{sec:sensitivity}
Following in general the procedure described in appendix~B of Ref.\,\cite{Gorbunov:2023jnx}, but considering now the results for the cross section of inelastic bremsstrahlung~\eqref{eq:final} obtained in this work, we estimate the sensitivity of the SHiP experiment to the visible decays of dark photons. The resulting 95\% CL exclusion limits in dark photon parameter space $(m_{\gamma^\prime},\epsilon)$ are presented in Fig.\,\ref{fig:sensitivity}
%============================= Fig. 3 ================================
\begin{figure}[t]
\begin{center}
\includegraphics[width=0.6\textwidth]{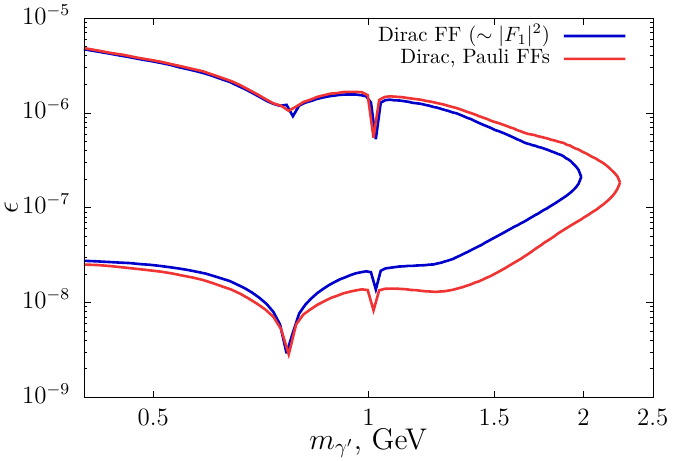}
\vspace{-3mm}
\caption{Contours in the dark photon parameter space for regions excluded at 95\%\,CL if at least three events of visible dark photon decay to charged pairs of SM particles are not observed in the SHiP experiment. Dark photons are assumed to be produced exclusively via inelastic bremsstrahlung. Red line shows the sensitivity taking into account both Dirac and Pauli form factors, whereas blue line corresponds to the conservative estimate only with Dirac form factor.}
\end{center}
\labelf{fig:sensitivity}
\vspace{-5mm}
\end{figure}
%============================= Fig. 3 ================================
in red. For comparison, the blue line shows the conservative sensitivity 95\%\,CL estimate that includes only the contribution of Dirac electromagnetic form factor of proton. The increase in total cross section~\eqref{eq:total} results in higher sensitivity and in the extension of the region where SHiP can search for dark photons. Thus it is important to update similar sensitivity curves for T2K and DUNE experiments, which is a subject for future work.

\section{Conclusions}
\label{sec:conclude}
To sum up, we have calculated new contributions to the dark photon production via inelastic proton bremsstrahlung. Being  proportional to the Pauli electromagnetic proton form factor, they are non-negligible at least for dark photon masses  $m_{\gamma^\prime}=0.9\,-\,1.8$\,GeV. They must be taken into account while estimating the sensitivities of future projects aimed at dark photon searches, and we illustrate its expected impacts by refining the sensitivity of the SHiP experiment.

\section*{Acknowledgements}
\label{sec:acknowledgement}
We thank Y.~H.~Lin and S.~Dubnicka for providing us with the tables with electromagnetic proton form factors. We would also like to thank S.~Godunov, O.~Teryaev, V.~Troitsky and M.~Vysotsky for helpful discussions. EK is grateful to M.~Matveev and E.~Kuzminskii for interesting and stimulating comments.  

\section*{Funding}
\label{sec:funding}
The work of EK is supported by the grant of the Foundation for the Advancement of Theoretical Physics and Mathematics “BASIS” no. 21-2-10-37-1.

\section*{Conflict of interest}
\label{sec:conflict}
The authors declare no conflicts of interest.

%\nocite{*}
\bibliographystyle{pepan}
\bibliography{pepan_biblio}

\end{document}